# Constraints on the general solutions of Einstein cosmological equations by Hubble parameter times cosmic age: a historical perspective

JULIO A. GONZALO[1] AND MANUEL ALFONSECA[2]

[1] Escuela Politécnica Superior, Universidad San Pablo CEU, Montepríncipe, Bohadilla del Monte, 28668 Madrid, Spain. Departamento de Física de Materiales, Universidad Autónoma de Madrid, 28049, Madrid, Spain.
[2] Escuela Politécnica Superior, Universidad San Pablo CEU, Universidad Autónoma de Madrid, 28049, Madrid, Spain.

Corresponding author : Manuel Alfonseca, email: manuel.alfonseca@uam.es . Phone: +34914972278. Fax:+34914972235

Keywords: Cosmic dynamical equations / Flat cosmic solutions / Open cosmic solutions / Intermediate cosmic solutions

**Abstract**

In a historical perspective, compact solutions of Einstein's equations, including the cosmological constant and the curvature terms, are obtained, starting from two recent observational estimates of the Hubble's parameter ($H_0$) and the "age" of the universe ($t_0$). Cosmological implications for $\Lambda$CDM ($\Lambda$ Cold Dark Matter), KOFL (k Open Friedman-Lemaitre), plus two mixed solutions are investigated, under the constraints imposed by the relatively narrow current uncertainties. Quantitative results obtained for the KOFL case seem to be compatible with matter density and the highest observed red-shifts from distant galaxies, while those obtained for the $\Lambda$CDM may be more difficult to reconcile.

**1. Introduction: relevance of the dimensionless product $H_0 t_0$**

In the late 70's, large uncertainties surrounded the numerical estimates of Hubble's ratio ($H_0 = \dot{R}_0 / R_0$) and the "age" of the universe. The uncertainty in the density parameter ($\Omega = \rho_0 / \rho_{co}$) giving the ratio between the current density and the critical density, $\rho_{co} = 3H_0^2 / 8\pi G$, was also very large.

An attempt was made by Beatriz Tinsley (Tinsley 1977), to explore what type of universe we live in, with the scarce information then available. She pointed out then that the Universe was most likely open, assuming a zero cosmological constant ($\Lambda = 0$) in Einstein's cosmological equations. Further, she noted that the dimensionless product $H_0 t_0$, which requires only local data, can be especially advantageous to characterize the cosmic equation.

At that time, many cosmologists stated their preference for a closed universe. They hoped that the amount of ordinary matter in the universe would be proven large enough to close the universe, but not by much, since otherwise the expansion would have been reversed by this time, something that obviously has not happened.

In the nineties, at a Summer Course on Astrophysical Cosmology (Gonzalo et al. 1995) which took place in El Escorial, Spain, a number of distinguished experts were present, including Ralph Alpher, John C. Mather, George F. Smoot, Hans Elsässer and Stanley L. Jaki. The general consensus in the years before that time was that $H_0$ should be somewhere between 50 and 100 Km/s/Mpc, and $t_0$ somewhere between 10 and 20 Gyrs.

The same year, the Harvard's group (Riess et al. 1995) offered a new, far better estimation, obtained by analyzing type Ia supernovae in far galaxies. $H_0$ was approximated at 67 Km/s/Mpc and $t_0$ at 13.7 Gyrs.

Today the old uncertainties have come down sharply. New accurate estimates have been given for $H_0$ and $t_0$. In particular, those obtained from the Wilkinson Microwave Anisotropy Prove (WMAP) and published on December 2012 (Komatsu et al. 2011) (Bennett et al. 2012) give:

$$H_0 = 69.3 \pm 1.8 \, \text{km s}^{-1} \text{Mpc}^{-1} = 2.246 \times 10^{-18} \, \text{s}^{-1} \qquad (1)$$

$$t_0 = 13.77 \pm 0.11 \, \text{Gyr.} = 4.345 \times 10^{17} \, s. \qquad (2)$$

while those published on March 2013 as a result of the analysis of the Planck telescope data (NASA 2013) (Ade et al. 2013) give:

$$H_0 = 67.15 \pm 1.2 \, \text{km s}^{-1} \text{Mpc}^{-1} = 2.176 \times 10^{-18} \, \text{s}^{-1} \qquad (3)$$

$$t_0 = 13.798 \pm 0.037 \, \text{Gyr.} = 4.354 \times 10^{17} \, s. \qquad (4)$$

Curiously enough, the latest estimation for H0 goes back to the value given in 1995 by Harvard's group. We have decided to consider both alternatives.

These data result in the dimensionless products $H_0 t_0 = 0.9759$ and $H_0 t_0 = 0.9476$ respectively, both smaller, but relatively close, to unity.

During the eighties, better and better measurements made clear that the amount of ordinary matter is much smaller (about 5%) than would be needed for a flat or a marginally closed universe. This, and some irregularities in the rotation of galaxies, brought to the conclusion that there must exist another kind of matter (dark matter) that would close the universe or at least make it flat. All the searches during the next thirty years failed to underpin dark matter, but the current estimations are quite smaller (about 27% of the critical value).

In 1998, a further analysis of type Ia supernovae by the Harvard group gave the (then) unexpected result that the universe is accelerating (Riess et al. 1998). This gave rise to the following consequences:

- The cosmological constant in Einstein's cosmological equation (see below) was resurrected. For several decades, its value had been assumed to be zero.
- A new mysterious entity (dark energy) was introduced to represent the effect of the cosmological constant. Its effect (currently unexplained) would be the same as a negative gravity, giving rise to the currently accelerated expansion of the universe.
- The flat model of the universe became the standard cosmological assumption. The amount of dark energy was estimated as precisely what was needed to make that model possible (about 68% of the total mass of the universe). The closed model has been abandoned, as it is incompatible with the acceleration. Research on the possibility of an open model with a non-zero curvature is now also neglected.

However, even at the time when the acceleration effect was discovered, other reasons apart from dark energy were suggested, which could explain at least a part of the effect. Brightness attenuation due to early substantially denser cosmic dust (Weinberg 2008), was offered as a possible explanation. Another one is proposed here:

Galaxies recede from the cosmic center of mass (assumed to coincide with the center of the expanding Cosmic Microwave Background Radiation sphere). A galaxy at a distance r=$R_0$ − R from the Milky Way galaxy (which moves at $v_0$≈2×10$^{-3}$c from the center of the CMBR sphere), recedes from us with a velocity v. These two variables, r and v, are characterized by the red-shift z of that galaxy in this way:

$$v = c \frac{(1+z)^2 - 1}{(1+z)^2 + 1} \quad (5)$$

$$r = R_0 \frac{z}{1+z} \quad (6)$$

This means that, for z<<1, both v/c and r/$R_0$ approach z. Here $R_0$ is the radius of the expanding observable universe, which is greater than the radius of the CMBR sphere, but close to it at present, since the CMBR is receding from the Milky Way at z≈1089, which corresponds (Weinberg 2008) to a velocity very close to c.

Figure 1 depicts $\log_{10}$(r/$R_0$) versus $\log_{10}$(v/c) as given respectively by equations (6) and (5), for z varying from 0.1 to 100. It shows that for z>1 (see marked points for z=0.5 to 10) the assumed proportionality between distance (r) and recession velocity (v) is no longer valid. Of course, for small values of z the proportionality is almost perfect. It should be noticed that observed values as large as z=10 have been reported for distant galaxies. The upward trend of distance (or equivalently magnitude) vs recession speed (velocity) in figure 1 mimics what is to be expected in an accelerated expansion, but is not related to a non-zero cosmological constant. So, it may explain, at least in part, the upward trend in magnitude vs red shift observed in the case of type Ia supernovae, for z>1 (the effect discovered in 1998).

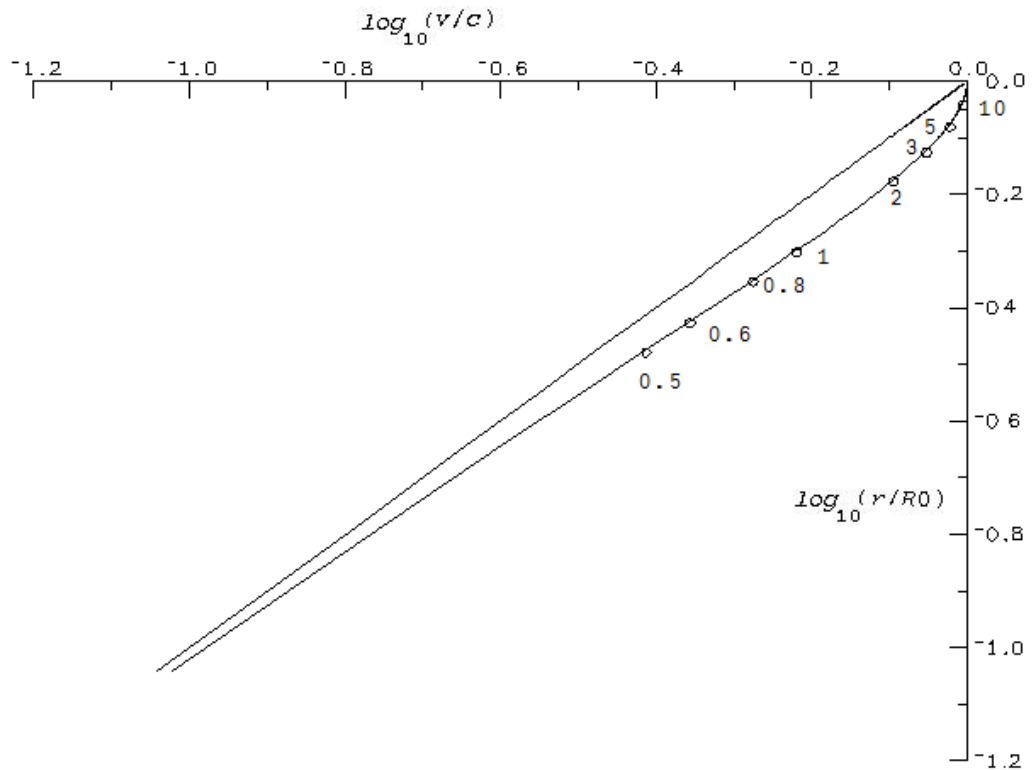

**Figure 1. Distance/velocity curve parametrized by redshift z**

## 2. Parametric solutions for Einstein's universe

Our starting point is the finite universe described by Einstein's cosmological equation (Einstein, 1923) which can be written as

$$\dot{R}^2 = \frac{2GM}{R} - kc^2 + \frac{\Lambda}{3}c^2 R^2, \qquad (7)$$

where $\dot{R}$ is the time derivative of the radius of the observable universe R, $G = 6.67 \times 10^{-11}$ IS units, M the finite mass of the observable universe, k the space-time curvature (k < 0 for an open universe), c = 3 x $10^8$ m/s the speed of light, and $\Lambda$ (cm$^{-2}$), Einstein cosmological constant, originally introduced by Einstein to counter gravitation.

Compact parametric solutions of Eq.(7) can be obtained easily in terms of *t(y)* and *R(y)*, where *y* is a cosmic parameter going from *y <<1* just after the singularity (*t = 0, R = 0*), to *y>>1* well away from it. Explicit analytical solutions can be obtained for (i) a **flat universe (k = 0, $\Lambda$ > 0)**, (not previously reported in compact parametric form, as far as we know); (ii) an **open universe (k < 0, $\Lambda$ = 0)**, equivalent to the well-known Friedman-Lemaître solution. There is also a third case, (iii) a **mixed universe (k < 0, $\Lambda$ > 0)**, where analytical solutions are not available, but the equations can be solved numerically.

In each case, from t(y) and R(y), $\dot{R}(y)$, etc, can be derived, as well as simple dimensionless expressions for H$_0$(y)t$_0$(y) and $\Omega_0(y)$, which will be shown useful to make direct quantitative comparisons with observational data, putting stringent constraints on the validity of the respective solutions.

### 2.1. Parametric solutions for a flat $\Lambda$CDM universe

With k = 0, $\Lambda$ > 0, Eq. (7) results in:

$$\int_0^t dt = \int_0^R \frac{R^{1/2}}{\left(2GM + \left(\frac{\Lambda}{3}c^2\right)R^3\right)^{1/2}} dR = \frac{2/3}{\left(\frac{\Lambda}{3}c^2\right)^{1/2}} \int_0^x \frac{1}{(a^2 + x^2)^{1/2}} dx \qquad (8)$$

where $x^2 \equiv \left(\frac{\Lambda}{3}c^2\right)R^3$, $a^2 \equiv \left(\frac{\Lambda}{3}c^2\right)R^3(\Lambda)$, which implies $R(\Lambda) = \left(2GM \big/ \frac{\Lambda}{3}c^2\right)^{1/3}$,

This integral can be solved analytically using the variable change $\frac{x}{a} = \sinh y$ and gives

$$t = \frac{2}{3}\left(\frac{\Lambda}{3}c^2\right)^{-\frac{1}{2}} y, \quad R = R(\Lambda)\sinh^{2/3} y, \quad \Lambda = \frac{4y^2}{3t^2 c^2} \qquad (9)$$

It can be seen that $\Lambda$, as a function of time, is completely determined by Einstein's equation in this model. Therefore, its current value ($\Lambda$0) depends exclusively from y0 and t0, or, in other words, from the current estimations of H0 and t0 (see Table I below).

We can proceed now to get the relevant cosmic parameters.

For the speed at which the cosmic radius is growing we get

$$\dot{R}(y) = \frac{dR/dy}{dt/dy} = R(\Lambda)\left(\frac{\Lambda}{3}c^2\right)^{1/2} \sinh^{-1/3} y \cosh y \qquad (10)$$

Hubble's parameter then becomes

$$H(y) = \frac{\dot{R}(y)}{R(y)} = \left(\frac{\Lambda}{3}c^2\right)^{1/2} \frac{\cosh y}{\sinh y} \quad (11)$$

and H(y)t(y), using Eqs.(11) and (9), results in

$$H(y)t(y) = \frac{2}{3} \frac{y}{\tanh y}, \quad (12)$$

which goes from H(0)t(0) = 2/3 to H(y)t(y) growing indefinitely for y >> 1.

The dimensionless density parameter $\Omega(y)$ becomes

$$\Omega(y) = \frac{\rho(y)}{\rho_c(y)} = \frac{M / \frac{4\pi}{3} R^3(y)}{3H^2(y)/8\pi G} = 1 - \tanh^2 y \quad (13)$$

Table I shows some of the results obtained for the two boundary conditions we have used: equations (1), (2) on the one hand, and (3), (4) on the other.

**Table I**

**Cosmic parameters for a flat (ΛCDM) universe. Left, $H_0$=69.3 km/seg/Mpc, $t_0$=13.77 Gyr, $\Lambda_0$= 1.2001e-52 m$^{-2}$, M0=2.5811e52 kg. Right, $H_0$=67.15 km/seg/Mpc, $t_0$=13.798 Gyr, $\Lambda_0$=1.0764e-52 m$^{-2}$, M0=2.9595e52 kg. In both cases, $T_0$=2.72548 K.**

| Radius | WMAP | | | | | Planck | | | | |
|---|---|---|---|---|---|---|---|---|---|---|
| | R (Mly) | y | t (My) | z | $\Omega_m$ | R (Mly) | y | t (My) | z | $\Omega_m$ |
| $R_0$ | 14110 | 1.2359 | 13770 | 0 | 0.287 | 14562 | 1.1729 | 13798 | 0 | 0.319 |
| $R_\Lambda$ | 10421 | 0.8814 | 9820 | 0.354 | 0.5 | 11310 | 0.8814 | 10369 | 0.2875 | 0.5 |
| $R_{Sch}$ | 4052 | 0.2401 | 2676 | 2.482 | 0.944 | 4646 | 0.2603 | 3063 | 2.134 | 0.935 |
| $R_{CMBR}$ | 12.82 | 4.3e-5 | 0.48 | 1099.7 | 1 | 13.23 | 4e-5 | 0.47 | 1099.7 | 1 |

In this case, the differences between the two sets of cosmic parameters are not too high. The time of the CMB radiation is nearest to the number usually given (370,000 years after the Big Bang): for both scenarios, the times computed are 470,000 and 480,000 years, respectively. But the value of z when the radius of the observable universe was equal to the Schwarzschild radius for the computed mass of the observable universe seems too low, as will be explained later. Such values would have the consequence that most of the far currently visible galaxies would have started to form when the whole universe was still an exploding black hole.

**2.2. Parametric solutions for an open KOFL universe**

Eq.(7), with k < 0, Λ = 0, results in:

$$\int_0^t dt = \int_0^R \frac{R^{1/2}}{\left(2GM + c^2|k|R\right)^{1/2}} dR = \frac{2}{c^3|k|^{3/2}} \int_0^x \frac{x^2}{\left(a^2 + x^2\right)^{1/2}} dx, \quad (14)$$

where $x^2 \equiv c^2|k|R$, $a^2 \equiv c^2|k|R_+$, which implies $R_+ = 2GM/|k|c^2$.

The integral can be solved analytically using the variable change $\frac{x}{a} = \sinh y$ and gives

$$t = \frac{R_+}{c|k|^{1/2}}[\sinh y \cosh y - y], \quad R = R_+ \sinh^2 y \tag{15}$$

We can proceed now to get the relevant cosmic parameters.

For the speed at which the cosmic radius is growing we get

$$\dot{R}(y) = \frac{dR(y)/dy}{dt(y)/dy} = |k|^{1/2} c \frac{1}{\tanh y} \tag{16}$$

Hubble's parameter then becomes

$$H(y) = \frac{\dot{R}(y)}{R(y)} = \frac{|k|^{1/2} c}{R_+} \frac{\cosh y}{\sinh^3 y} \tag{17}$$

and $H(y)t(y)$, using Eqs.(17) and (15), results in

$$H(y)t(y) = \frac{1}{\tanh^2 y} - \frac{y}{\tanh y \sinh^2 y}, \tag{18}$$

which goes from $H(0)t(0) = 2/3$ to $H(y)t(y) = 1$ for $y \gg 1$.

The dimensionless density parameter $\Omega(y) = \rho_m/\rho_{mc}$ (which in fact corresponds to $\Omega_m$) becomes

$$\Omega(y) = \frac{\rho(y)}{\rho_c(y)} = 1 - \tanh^2 y \tag{19}$$

This analytical expression is identical to the one derived for the open universe, although one must remember that the meaning of parameter y is different in both situations.

Table II shows some of the results obtained for the two boundary conditions we have used: equations (1), (2) on the one hand, and (3), (4) on the other. In this case, the differences between the two sets of cosmic parameters are higher than in the flat universe, specially as regards the Schwarzschild radius and time, which are over three times larger for a slightly smaller value of $H_0$. This means that this model is very sensitive to the actual value of $H_0$. In both cases, however, the corresponding value of z is higher than the currently observed farthest galaxy, which means that all of them were formed when the observable universe was not an exploding black hole. The time of the formation of the CMB radiation is significantly higher than in the flat case: 1,381,000 and 2,253,000 years, respectively.

**Table II**
**Cosmic parameters for an open (KOFL) universe, k=-1. Left, $H_0$=69.3 km/seg/Mpc, $t_0$=13.77 Gyr, M0=1.1479e51 kg. Right, $H_0$=67.15 km/seg/Mpc, $t_0$=13.798 Gyr, M0=3.585e51 kg. In both cases, $T_0$=2.72548 K.**

| Radius | WMAP | | | | | Planck | | | | |
|---|---|---|---|---|---|---|---|---|---|---|
| | R (Mly) | y | t (My) | z | $\Omega_m$ | R (Mly) | y | t (My) | z | $\Omega_m$ |
| $R_0$ | 14199 | 2.8797 | 13770 | 0 | 0.0125 | 14835 | 2.3384 | 13798 | 0 | 0.0365 |
| $R_+=R_{Sch}$ | 180.20 | 0.8814 | 96.0 | 77.8 | 0.5 | 562.78 | 0.8814 | 299.9 | 25.36 | 0.5 |
| $R_{CMBR}$ | 12.90 | 0.2645 | 2.253 | 1099.7 | 0.933 | 13.48 | 0.1541 | 1.381 | 1099.7 | 0.977 |

Table III compares some of the results obtained for the open (KOFL universe) with several values of the curvature k. It can be seen that the time computed for the formation of the CMB radiation does not depend on the value of k. The radius of the observable universe and the Schwarzschild radius, however, get quickly smaller when k goes near to zero. In fact, the first radius is smaller than the distance run by light since the CMB radiation formed, for all k≥0.9 in the first scenario, and for all k>0.9 in the second scenario.

**Table III**
**Results comparison for an open (KOFL) universe with different values of k. Left, $H_0$=69.3 km/seg/Mpc, $t_0$=13.77 Gyr, M0=1.1479e51 kg. Right, $H_0$=67.15 km/seg/Mpc, $t_0$=13.798 Gyr, M0=3.585e51 kg. In both cases, $T_0$=2.72548 K.**

| k | WMAP | | | | Planck | | | |
|---|---|---|---|---|---|---|---|---|
| | R (Mly) | $R_{Sch}$ (Mly) | $z_{Sch}$ | $t_{CMBR}$ | R (Mly) | $R_{Sch}$ (Mly) | $z_{Sch}$ | $t_{CMBR}$ |
| -1 | 14199 | 180.2 | 77.8 | 2.253 | 14835 | 562.8 | 25.36 | 1.381 |
| -0.9 | 13470 | 153.9 | 86.6 | 2.253 | 14074 | 480.5 | 28.29 | 1.381 |
| -0.75 | 12297 | 117.0 | 104.1 | 2.253 | 12848 | 365.5 | 34.15 | 1.381 |
| -0.5 | 10040 | 63.7 | 156.6 | 2.253 | 10490 | 199.0 | 51.72 | 1.381 |
| -0.1 | 4490 | 5.7 | 787.0 | 2.253 | 4691 | 17.8 | 262.6 | 1.381 |

## 2.3. Numerical solutions for a mixed universe

The numerical solutions for a mixed universe (**k < 0, Λ > 0**), can be easily obtained by solving the Einstein equation (7) for the appropriate value of the mass of the observable universe. As this value is unknown, this parameter must be adjusted by successive approximations. The Einstein equation cannot be solved numerically starting at t=0 (the Big Bang itself), because at that point there is a singularity, therefore we decided to start solving the equation at t=$t_{CMBR}$, which means that the initial condition (the time at which this phenomenon took place) must also be estimated. We did it by interpolating between the corresponding values for the flat universe solution and the open universe solution. Table IV shows the results for k=-0.5, Λ= $Λ_0$/2, M0=1.375.$10^{52}$/1.9114.$10^{52}$ kg and k=-0.75, Λ= $Λ_0$/4, M0=5.25.$10^{51}$/9.8.$10^{51}$ kg (using the value of $Λ_0$ for each scenario).

**Table IV**
**Cosmic parameters for a mixed universe. Left, $H_0$=69.3 km/seg/Mpc, $t_0$=13.77 Gyr. Right, $H_0$=67.15 km/seg/Mpc, $t_0$=13.798 Gyr. In both cases, $T_0$=2.72548 K.**

| Radius | WMAP | | | | Planck | | | |
|---|---|---|---|---|---|---|---|---|
| Λ=$Λ_0$/2, k=-0.5 | R (Mly) | t (My) | z | $Ω_m$ | R (Mly) | t (My) | z | $Ω_m$ |
| $R_0$ | 13833 | 13770 | 0 | 0.124 | 15409 | 13798 | 0 | 0.174 |
| $R_{Sch}$ | 1648 | 967.3 | 7.4 | 0.665 | 3000 | 1758.5 | 4.14 | 0.660 |
| $R_{CMBR}$ | 12.567 | 0.93 | 1099.7 | 0.996 | 14.0 | 0.92 | 1099.7 | 0.9977 |
| Radius Λ=$Λ_0$/4, k=-0.75 | R (Mly) | t (My) | z | $Ω_m$ | R (Mly) | t (My) | z | $Ω_m$ |
| $R_0$ | 13989 | 13770 | 0 | 0.0599 | 14940 | 13798 | 0 | 0.0993 |
| $R_{Sch}$ | 824 | 459.8 | 15.97 | 0.571 | 1538 | 858.2 | 8.71 | 0.571 |
| $R_{CMBR}$ | 12.709 | 1.15 | 1099.7 | 0.9886 | 13.57 | 1.15 | 1099.7 | 0.9934 |

Looking at table IV, it can be seen that only the left lower part is compatible with the formation of all visible galaxies after the observable universe stopped being an exploding black hole. The origin of the CMB radiation in these cases would have happened around one million years after the Big Bang.

**3. Constraints on the time dependence of the density parameter Ωm**

Figure 2 shows Ω(y) vs H(y)t(y) for a **flat** universe, Eq. 13, and an **open** universe, Eq.(19), using the Planck data. Note that the current situation ($H_0 t_0$) is represented in both models by the same abscissa (the dotted line), while the half-density parameter Ω(y)=1/2, signaled by separate arrows for both models, is substantially different (see Tables I and II). H(y)t(y)<2/3 corresponds to closed universes, currently discarded. For the open universe (the dark blue line), 2/3<H(y)t(y)<1; for the flat universe (the red line), 2/3<H(y)t(y)<∞. The section tinted in light blue would correspond to the whole range of different mixed cases.

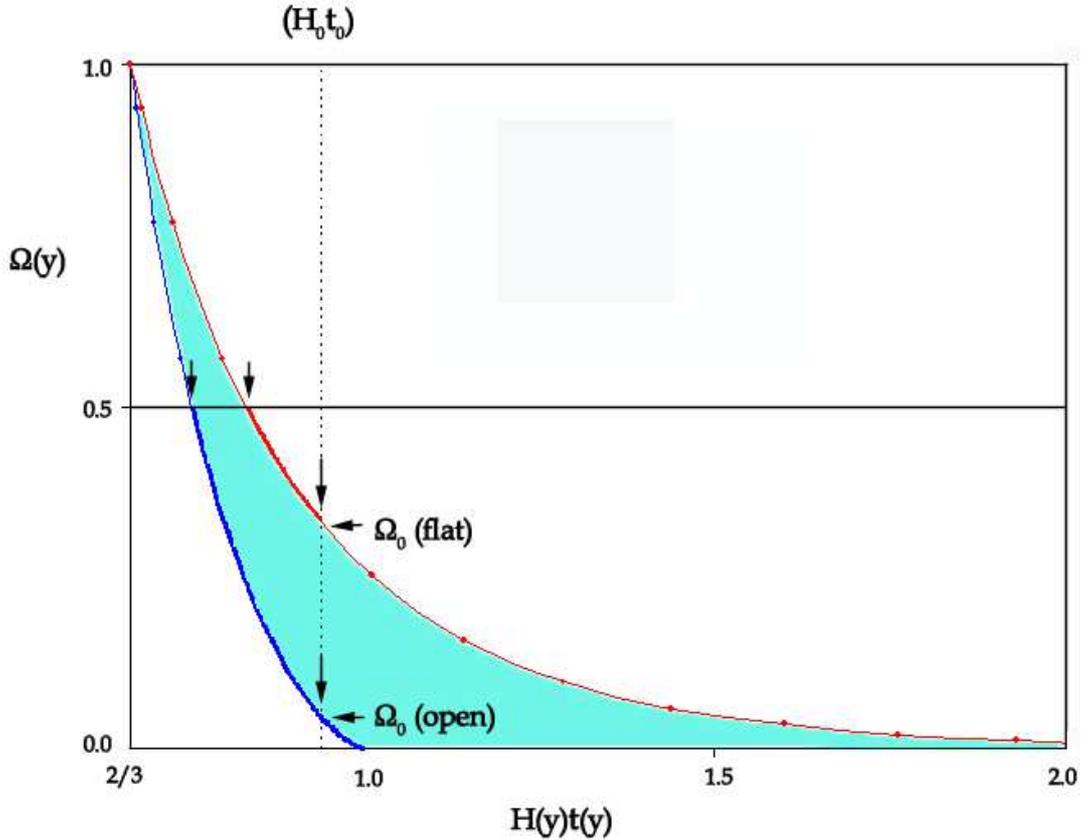

**Figure 2 Matter density parameter Ω(y) vs. dimensionless cosmic parameter H(y)t(y) = Hubble's ratio x time for an open (OFLM) and a flat (ΛCDM) universe**

Equation (7) can be rewritten (Weinberg 2008) as:

$$1 = \Omega_m + \Omega_k + \Omega_\Lambda \qquad (20)$$

where $\Omega_m$ is the time-dependent mass density parameter (actually mater mass plus radiation mass, although at present radiation density is much less than matter density), $\Omega_k$ is the time-dependent space-time curvature energy density (which equals zero if k=0), and $\Omega_\Lambda$ is the time-dependent energy density associated to Λ (which equals zero if Λ=0). This relationship holds at present and at any time since the big bang.

Figure 3a, for the flat (OFLM) model (ΛCDM) model, displays the evolution of Ω= $\Omega_m$+ $\Omega_r$ and $\Omega_\Lambda$ (due to the cosmological constant or "dark energy") as a function of z (obviously related to T(K), t(gyrs), y(dimensionless), from the cosmic microwave background radiation to z=0. The dotted lines correspond to $t_{Sch}$ (3.063 Gyrs) and $t_0$ (13.798 Gyrs). Figure 3b displays the evolution of Ω= $\Omega_m$+ $\Omega_r$ and $\Omega_k$ as a function of z, for the open (OFLM) model, where the dotted lines correspond to $t_{Sch}$ (0.2999 Gyrs) and $t_0$ (13.798 Gyrs). In both cases, the Planck data are used.

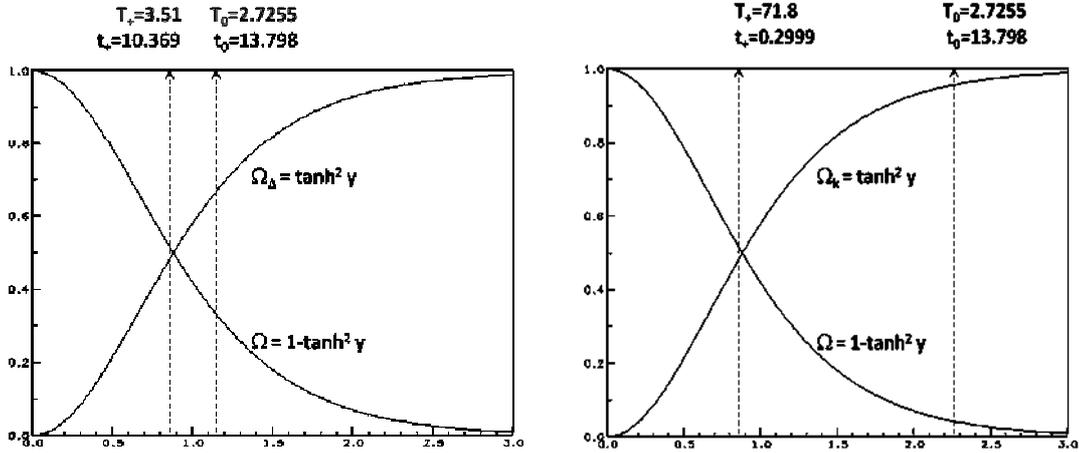

**Figure 3 a) Ω and $\Omega_\Lambda$ vs z for a flat (ΛCDM) model. b) Ω and $\Omega_k$ vs z for an open (KOFL) model**

When the radius of the universe grew beyond the Schwarzschild radius $R_{Sch} = 2GM/c^2$ (this happened long after baryon formation, nucleo-synthesis and atom formation, and the decoupling of the cosmic microwave background radiation) the universe was already transparent. We can assume that galaxies did not form before that time (when the behaviour of the universe would have been similar to that of an exploding black hole), but started forming soon after. Sometime later, when galaxies and stars were fully formed, they began to emit red-shifted light, which is now arriving to us. At present:

$$1 = \Omega_{m0} + \Omega_{k0} + \Omega_{\Lambda 0} \qquad (21)$$

And averaging from the time of galaxy formation to the present we have:

$$1 = <\Omega_m> + <\Omega_k> + <\Omega_\Lambda> \qquad (22)$$

This is illustrated in figure 4, which has been drawn using the data in equations (3) and (4), for the flat ΛCDM and for the open KOFL universe. A quantitative discussion will take place in section 4 for the cases corresponding to equations (1), (2) and (3), (4). It must be noted that in Figure 4 the respective values of $\Omega_m$ and $\Omega_\Lambda$ /$\Omega_k$ are time-dependent from the Big Bang ($\Omega_m$=1) to time growing indefinitely ($\Omega_m$→0) as required by equation (13).

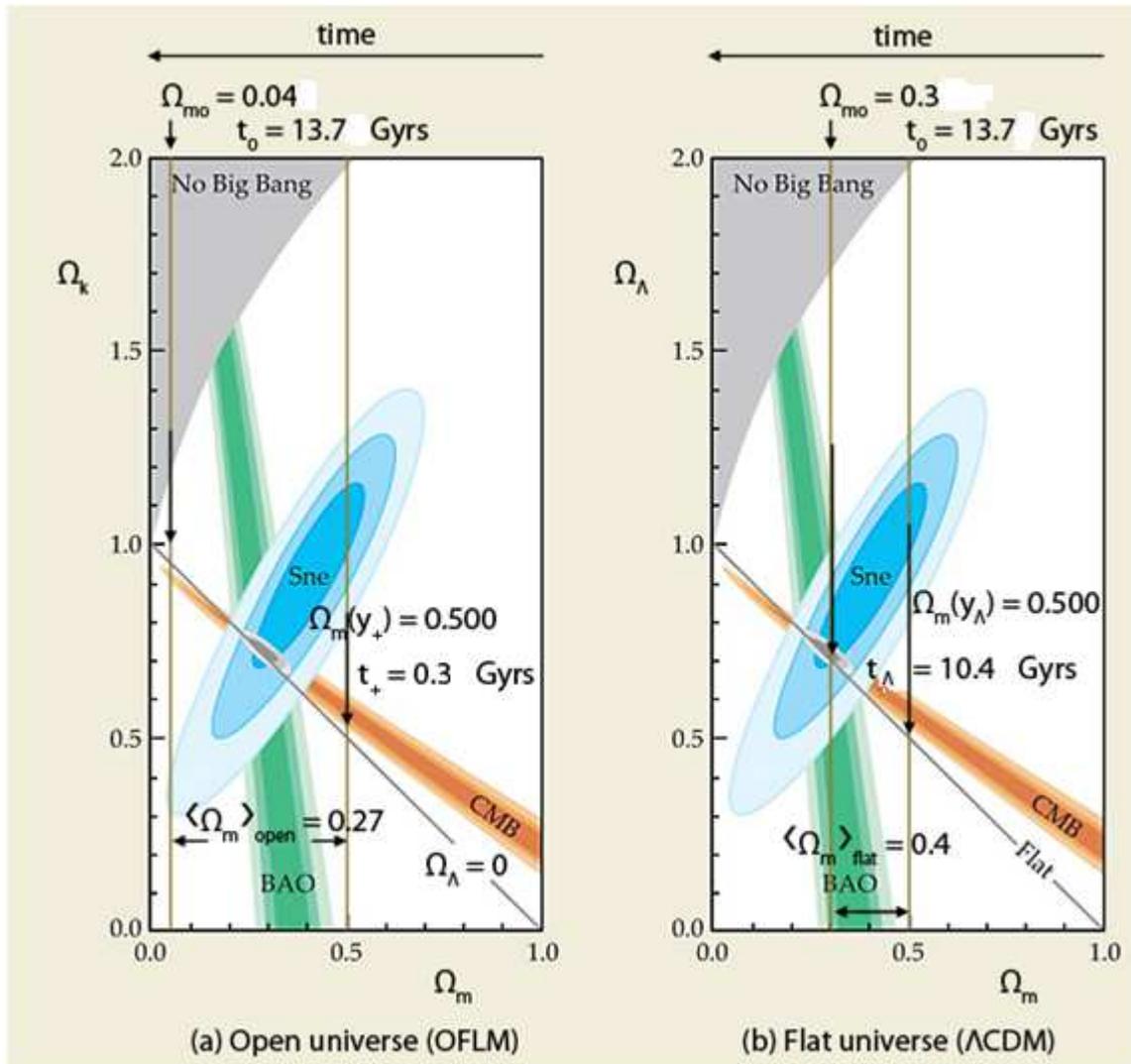

**Fig.4** Time evolution of (a) $\Omega_k$ vs. $\Omega_m$ for an open, and (b) $\Omega_\Lambda$ vs. $\Omega_m$ for a flat universe. Confidence contours from the cosmic microwave background (CMB) and other constraints are shown. (See **Physics Today**, Dec 2011, pp 14-17)

In summary, we have shown that, as anticipated by Beatriz Tinsley in the late 70's, a judicious use of the dimensionless product $H(y)t(y)$ with properly compact solutions of Einstein's cosmological equation, lead to a possible discrimination between the two main cosmological models discussed here.

### 4. Quantitative comparisons between flat ΛCDM, open KOFL and mixed solutions

Tables V and VI show the values predicted for several cosmic quantities for both the flat ΛCDM and the open KOFL solutions in two different cases:

- $H_0$=69.3 km/seg/Mpc, $t_0$=13.77 Gyr and
- $H_0$=67.15 km/seg/Mpc, $t_0$=13.798 Gyr

In both cases, $y_0$ has been determined from the dimensionless value of $H_0 t_0$, and the remaining values have been obtained thus:

- $R_0 = c/H_0$
- $R_{Sch} = R[y_{Sch} = \sinh^{-1}(1)]$

- $T_{Sch}=R_0T_0/R_{Sch}$
- $z_{Sch}=T_{Sch}/T_0 - 1$
- $\Omega_{m0}=1-\tanh^2(y_0)$
- $<\Omega_m>=(\Omega_{m0}+\Omega_{Sch})/2$
- $<\Omega_x>=1-<\Omega_m>$ with $x=\Lambda,k$

Two additional mixed cases, computed numerically, have been added to the tables.

**Table V**
**Cosmic parameters for a flat (ΛCDM), mixed and open (KOFL) universe with WMAP results:**
$H_0$=69.3 km/seg/Mpc, $t_0$=13.77 Gyr, $T_0$=2.726 K

| Model | $y_0$ | $M_u$ $10^{51}$ kg | $R_0$ Mly | $T_{Sch}$ K | $z_{Sch}$ | $\Omega_{m0}$ | $<\Omega_m>$ | $<\Omega_x>$ |
|---|---|---|---|---|---|---|---|---|
| ΛCDM (Λ$_0$>0) | 1.2359 | 25.81 | 14110 | 9.49 | 2.482 | 0.287 | 0.616 | 0.384 |
| KOFL (k=-1) | 2.8797 | 1.148 | 14199 | 214.8 | 77.8 | 0.0125 | 0.256 | 0.744 |
| Mixed (Λ$_0$/2, k=-0.5) | | 13.75 | 13833 | 22.87 | 7.39 | 0.124 | 0.394 | 0.606 |
| Mixed (Λ$_0$/4, k=-0.75) | | 5.25 | 13989 | 46.26 | 15.97 | 0.0599 | 0.315 | 0.685 |

**Table VI**
**Cosmic parameters for a flat (ΛCDM), mixed and open (KOFL) universe with Plank results:**
$H_0$=67.15 km/seg/Mpc, $t_0$=13.798 Gyr, $T_0$=2.726 K

| Model | $y_0$ | $M_u$ $10^{51}$ kg | $R_0$ Mly | $T_{Sch}$ K | $z_{Sch}$ | $\Omega_{m0}$ | $<\Omega_m>$ | $<\Omega_x>$ |
|---|---|---|---|---|---|---|---|---|
| ΛCDM (Λ>0) | 1.1729 | 29.60 | 14562 | 8.54 | 2.134 | 0.319 | 0.627 | 0.373 |
| KOFL (k=-1) | 2.3384 | 3.585 | 14835 | 71.8 | 25.36 | 0.0365 | 0.268 | 0.732 |
| Mixed (Λ$_0$/2, k=-0.5) | | 19.114 | 15409 | 14.0 | 4.14 | 0.174 | 0.417 | 0.583 |
| Mixed (Λ$_0$/4, k=-0.75) | | 9.8 | 14940 | 26.47 | 8.71 | 0.0993 | 0.335 | 0.665 |

It can be seen that the values for the Schwarzschild red-shift and the densities for the open (KOFL) universe are in better agreement with present observational expectations, especially in table VI. In both tables, the flat universe would be untenable, since the maximum observable red-shift for galaxies ($z_{Sch}$) would appear to be significantly less than $z \approx 10$, currently observed.

## 5. Concluding remarks

We have shown that, as pointed out long ago by Beatriz Tinsley, using the dimensionless product $H(y)t(y)$ in conjunction with the dimensionless density ratio $\Omega_m(y)$, imposes stringent constraints on the solutions of Einstein's cosmological equation, pointing to a better understanding of the dark matter/dark energy question. In this respect we note that ignoring the time dependence of $\Omega_m(y)$ and $H(y)t(y)$ is clearly misleading.

We have reached the following additional conclusions:

- If one assumes the flat **ΛCDM** model, the value of the cosmological constant is completely determined by Einstein's equation and its current value depends only on the estimations of the values of $H_0$ and $t_0$.
- The time usually associated to the formation of the CMB radiation (379,000 years after the Big Bang) was computed assuming that the **ΛCDM** model is the correct one and using a slightly smaller decoupling temperature (about 2970 K and z=1088.7, which we have approximated to 3000 K and z=1099.7). This time ($t_{CMBR}$) depends on the model used and is quite sensible both to the universe model and to the different combinations of values of $H_0$ and $t_0$. Its value for a **KOFL** model is much larger (over one million years).

- If we start from the assumption that galaxies probably started forming after the observable universe radius exceeded the Schwartzschild radius, our tests show that, in a flat universe, many of the oldest galaxies would have been formed when the universe was still an exploding black hole. An open universe, on the other hand, gives an universe where this does not happen for $|k|>0.9$. A mixed universe would also be compatible with this for smaller values of $|k|$ and $\Lambda_0>0$.

When the measurements of the apparent magnitudes of type Ia supernovae, indicative of an accelerated expansion of the universe, were reported by S.Perlmutter (Schwarzchild 2011), he signaled the possibility that part of the effect could be explained by a possible dimming by intervening dust. Later, this well-founded concern was discarded and a flat universe model has been generally assumed. However, taking into account that at early times the matter mass density was significantly higher than it is now, and so was presumably the cosmic dust density, the case for systematic corrections of the apparent magnitudes of type Ia supernovae is still worthy of consideration, together with the fact that an upwards trend at higher red-shifts in the Hubble plot of $log_{10}(r/R_0)$ versus $log_{10}(v/c)$, as shown in figure 1, could be interpreted as an accelerated expansion, completely unrelated to a non-zero cosmological constant. In this case, an open universe model could be compatible with the apparent acceleration of the universe.

**Acknowledgements**

We would like to thank D.N. Spergel for providing links to the complete WMAP data.

We are indebted to Ralph A. Alpher and Stanley L. Jaki (no longer with us) for many fruitful conversations and protracted correspondence on cosmological matters.

We are also very grateful to Manuel de la Pascua, Carmen Aragó, and Manuel I. Marqués for their support and valuable help in preparing the manuscript.

And to Manuel M. Carreira S.J., José L. Sánchez Gómez, Ginés Lifante, Manuel Tello, and Antonio Alfonso Faus for many helpful conversations.